\documentclass[galaxies,article,accept,moreauthors,pdftex,10pt,a4paper]{mdpi}
\firstpage{1}
\articlenumber{38}
\doinum{10.3390/galaxies4040038}
\pubvolume{4}
\pubyear{2016}
\copyrightyear{2016}
\externaleditor{{Academic Editors: Jose L. G\`omez, Alan P. Marscher and Svetlana G. Jorstad}}
\history{{Received: 13 July 2016; Accepted: 22 September 2016; Published: date}}

\usepackage{microtype}
\usepackage{soul}

\usepackage{subfigure}
\makeatletter
\renewcommand{\@thesubfigure}{\normalsize(\textbf{\alph{subfigure}})}
\makeatother
\newcommand\myurl[1]{\changeurlcolor{black}\url{#1}\changeurlcolor{blue}}
\makeatletter
\g@addto@macro{\UrlBreaks}{\UrlOrds}
\makeatother

\Title{Microscopic Processes in Global Relativistic Jets Containing Helical Magnetic Fields}

\Author{{Ken-Ichi Nishikawa} $^{1,}$*, Yosuke Mizuno $^{2}$,  Jacek Niemiec $^{3}$, Oleh Kobzar $^{3}$,
 Martin Pohl $^{4,5}$, Jose L. G\'{o}mez $^{6}$, Ioana Du\c{t}an $^{7}$, Asaf Pe'er $^{8}$,
 Jacob Trier Frederiksen $^{9}$,  \AA ke Nordlund $^{9}$, Athina Meli $^{10}$, Helene Sol $^{11}$,
 Philip E. Hardee $^{12}$
 \mbox{and Dieter. H. Hartmann $^{13}$ }}

\address{%
$^{1}$ \quad  Department of Physics, University of Alabama in Huntsville, ZP12, Huntsville, AL 35899, USA\\
$^{2}$ \quad Institute for Theoretical Physics, Goethe University,  Frankfurt am Main {D-60438}, Germany; mizuno@th.physik.uni-frankfurt.de\\
$^{3}$ \quad Institute of Nuclear Physics PAN, ul. Radzikowskiego
152,  Krak\'{o}w 31-342, Poland; Jacek.Niemiec@ifj.edu.pl~(J.N.); oleh.kobzar@ifj.edu.pl{(O.K.)}\\
$^{4}$ \quad  Institut fur Physik und Astronomie, Universit\"{a}t Potsdam,  Potsdam-Golm {14476}, Germany; pohlmadq@gmail.com\\
$^{5}$ \quad DESY, Platanenallee 6,  Zeuthen {15738}, Germany; pohlmadq@gmail.com\\
$^{6}$ \quad Instituto de Astrof\'{i}sica de Andaluc\'{i}a, CSIC, Apartado 3004, Granada 18080, Spain; jlgomez@iaa.csic.es\\
$^{7}$ \quad Institute of Space Science, Atomistilor 409, Bucharest-Magurele RO-077125, Romania; ioana.dutan@gmail.com\\
$^{8}$ \quad Physics Department, University College Cork, Cork T12 YN60, Ireland; a.peer@ucc.ie\\
$^{9}$ \quad Niels Bohr Institute, University of Copenhagen, Blegdamsvej 17, Copenhagen DK-2100, Denmark; trier@nbi.ku.dk (J.T.F.), aake@nbi.dk (A.N.)\\
$^{10}$\quad Department  of Physics and Astronomy, University of Gent, Proeftuinstraat 86,
Gent B-9000, Belgium; ameli@ulg.ac.be\\
$^{11}$\quad LUTH, Observatore de Paris-Meudon, 5 place Jules
Jansen, Meudon Cedex 92195, France; helene.sol@obspm.fr\\
$^{12}$\quad Department of Physics and Astronomy, The University
of Alabama, Tuscaloosa, AL 35487, USA; pehardee@gmail.com\\
$^{13}$\quad Department of Physics and Astronomy, Clemson University, Clemson, SC 29634, USA; hdieter@g.clemson.edu}

\corres{Correspondence: ken-ichi.nishikawa@nasa.gov; Tel.: +1-256-824-2593}

\abstract{In the study of relativistic jets one of the key open questions is their interaction with the environment on the microscopic level. Here, we study the initial evolution of both electron$-$proton ($e^{-}-p^{+}$) and electron$-$positron ($e^{\pm}$) relativistic jets containing helical magnetic fields, focusing on their interaction with an ambient plasma. We have performed simulations of ``global'' jets containing helical magnetic fields in order to examine how helical magnetic fields affect  kinetic instabilities such as the Weibel instability,  the kinetic Kelvin-Helmholtz instability (kKHI)  and the Mushroom instability (MI). In our initial simulation study these kinetic instabilities are suppressed  and new types of instabilities can grow.  In the $e^{-}-p^{+}$ jet simulation  a recollimation-like instability occurs and jet electrons are strongly perturbed. In the $e^{\pm}$ jet simulation a recollimation-like instability occurs at early times followed by a kinetic instability and the general structure is similar to a simulation without helical magnetic field. Simulations using much larger systems are required in order to thoroughly follow the evolution of global jets containing  helical magnetic fields.
}


\begin{document}



\section{Introduction}

Relativistic jets are collimated plasma outflows associated with active galactic nuclei (AGNs), gamma-ray bursts
(GRBs), and pulsars. Among these astrophysical systems, blazars and GRB jets produce the most luminous phenomena in the universe
~{(}e.g., \cite{peer14} {)}.  Despite extensive observational and theoretical investigations (including simulation studies), our understanding of their formation, 
interaction, and evolution in the ambient plasma
---and consequently their observable properties, such~as  time-dependent flux and polarity
---remains quite limited. One of the key open questions in the study of relativistic jets is how they interact with the immediate plasma environment on the microscopic scale. In particular,  we wish to examine how relativistic jets  containing helical magnetic fields evolve under the influence of kinetic and {MHD}-like
~instabilities that occur within and at the jet boundaries, with consequences such as flares due to reconnection.

Jet outflows are commonly thought to be dynamically hot (relativistic) magnetized plasma flows
launched, accelerated, and collimated in regions where Poynting flux dominates over particle (matter) flux
\cite{blanz77,mck14}.
This scenario involves a helical large-scale magnetic field structure in some AGN jets, which~provides a unique
signature in the form of observed asymmetries across the jet width, particularly in the polarization \cite{laing81,
aloy00,Clausen11}.

Large-scale
~ordered magnetic fields have been invoked to explain the launching, acceleration, and collimation of relativistic  jets from the central nuclear region of an active galaxy \cite{meier08}, and from coalescing and merging compact objects (neutron stars and black holes)
; e.g., \cite{piran04}. The magnetic field structure and
particle composition of the jets are still  not well constrained observationally. Circular~polarization (CP; measured as Stokes parameter V) in the radio continuum emission from AGN jets provides  a~powerful diagnostic for {the deduction of}
~magnetic structure and particle composition, because
{---}unlike linear polarization (LP)
{---}CP is expected to remain almost completely unmodified by external screens
(e.g.,~\cite{OS13}).

Jet particle composition has remained an unresolved issue ever since the discovery of jets. The~two~main candidates are a ``normal'' plasma consisting of relativistic electrons and protons (an~$e^{-}$
{--}$p^{+}$ jet),
and a ``pair'' plasma consisting only of relativistic electrons and positrons (an $e^{\pm}$~jet)~\cite{wardle98}.
The~detection of  circular polarization from the violently variable quasar 3C\, 279 at several epochs, using the Very Long Baseline Array (VLBA) at 15 GHz, has been used by Wardle et al. \cite{wardle98} to argue that the circular polarization is produced by Faraday conversion of linear to circular polarization in the jet plasma.  This conversion requires that the energy distribution of the radiating particles extends down to $\gamma_{\min} \ll 100$, and that should imply an  $e^{\pm}$ jet.


Over the past few years, we have been using a fully self-consistent relativistic particle-in-cell (RPIC) simulation
method to investigate collisionless shocks, and the kinetic Kelvin--Helmholtz instability
(kKHI) at relativistic jet--sheath shear boundaries, and to calculate the resulting synthetic emission spectra.
The RPIC code used in these studies is a modified version of the TRISTAN code~\cite{buneman93},  parallelized with MPI and utilized for various research projects
~(e.g., \cite{niem08,nishi09,nishi16}).
To date, RPIC simulations of the kKHI have been performed in slab \cite{Alves12,Alves14,Alves15,Gris13a,Gris13b,
nishi13a,nishi13b,nishi14a,liang13a,liang13b}
and  cylindrical  geometries using periodic boundary conditions \cite{Alves10,nishi14b}.
Previously, full-scale shock simulations have not incorporated velocity shear interactions at the jet boundary with the ambient plasma (interstellar medium)
~(e.g.,~\cite{nishi09}), and~global shock simulations including velocity shear interactions performed to date used only very small simulation boxes
 \cite{nishi03,nishi05,ng06}.
Recently, we performed ``global'' jet simulations involving {the} injection of a~cylindrical unmagnetized jet into an ambient plasma in order to investigate shock (Weibel instability) and velocity shear instabilities (kKHI and {Mushroom instability (MI))} simultaneously \cite{nishi16}. Here we report preliminary results of our new studies of global relativistic jets containing helical magnetic~fields.


\section{Global Jet Simulations}



Jets generated  from black holes and injected into the ambient interstellar medium contain magnetic fields which are thought to be helical. Therefore, we perform global simulations of jets containing helical magnetic fields injected
into an ambient medium
~(e.g., \cite{mck14}).
The key issue we investigate is how the helical magnetic fields affect the growth of the kKHI, the MI, and the Weibel instability. It is known from RMHD
~simulations that jets containing helical magnetic fields develop 
~{kink} instability
~{(}e.g.,~\cite{mizuno14,singh16}{)}.  Recently, it has been demonstrated by Markidis et al. \cite{markid14} that a kinked jet structure leads to the occurrence of a secondary reconnection.

\vspace*{-0.0cm}
\subsection{Helical Magnetic Field Structure}
\vspace*{-0.0cm}

In our simulations, cylindrical jets are injected with a helical magnetic field  (see Figure \ref{hmfg1}a)
\cite{markid14}
implemented like that in RMHD simulations performed by Mizuno et al. \cite{mizuno15}. 
 Our toroidal magnetic field structure is similar to a simple
screw--pinch configuration adopted by \cite{markid14} with the radially dependent axial magnetic field.
However, our simulations use Cartesian coordinates.
Since $\alpha =1$, Equations (9)--(11) from \cite{mizuno15} are reduced to Equation (1), and the magnetic field  takes the form:
\begin{eqnarray}
B_{x} = \frac{B_{0}}{[1 + (r/a)^2]}, \,  \, \, \, \, \,  B_{\phi} =  \frac{(r/a)B_{0}}{[1 + (r/a)^2]}
\end{eqnarray}

The toroidal magnetic field is created by a current $+J_{x}(y, z)$ in the positive $x$-direction, so that
defined in Cartesian coordinates:
\begin{eqnarray}
B_{y}(y, z) =  \frac{((z-z_{\rm jc})/a)B_{0}}{[1 + (r/a)^2]}, \, \, \,\,  \,  \,
B_{z}(y, z) =  -\frac{((y-y_{\rm jc})/a)B_{0}}{[1 + (r/a)^2]}.
\end{eqnarray}

Here  $a$ is  the characteristic  length-scale of the helical magnetic field, $(y_{\rm jc},\, z_{\rm jc})$ is the center of
the jet, and  $r = \sqrt{(y-y_{\rm jc})^2+(z-z_{\rm jc})^2}$. The chosen helicity is defined through Equation (2),
which is left-handed polarity with positive $B_0$. At the jet orifice, we  implement  the helical magnetic field without the motional electric fields. This corresponds to a toroidal magnetic field generated self-consistently by jet particles moving along{ the} $+x$-direction.

\vspace{-0.0cm}
\subsection{Currents and Fields in Helically Magnetized RPIC Jets}
\vspace{-0.0cm}

\begin{figure}[H]
\centering
\includegraphics[width=55mm]{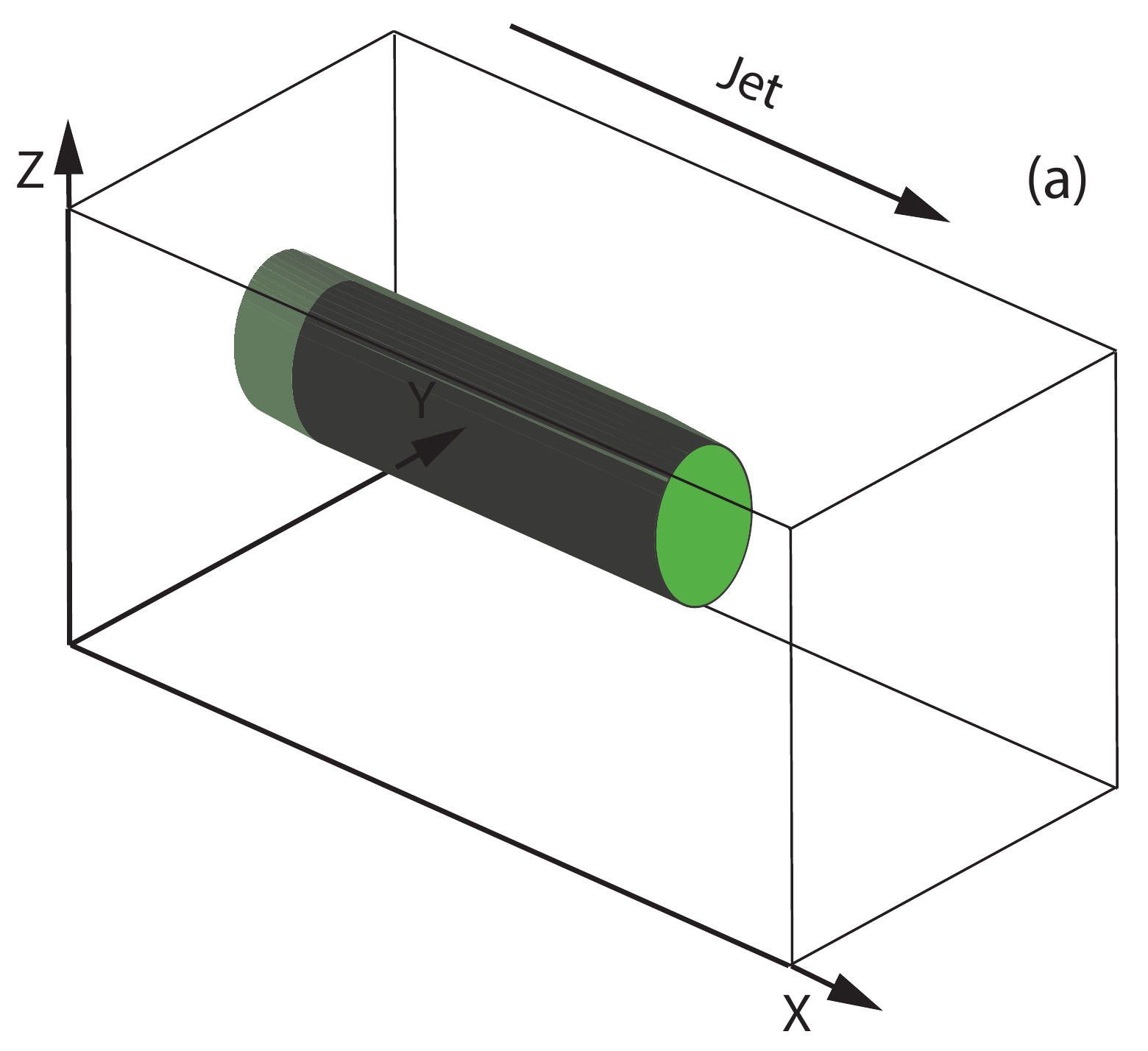}
\hspace{1.5cm}
\includegraphics[width=50mm]{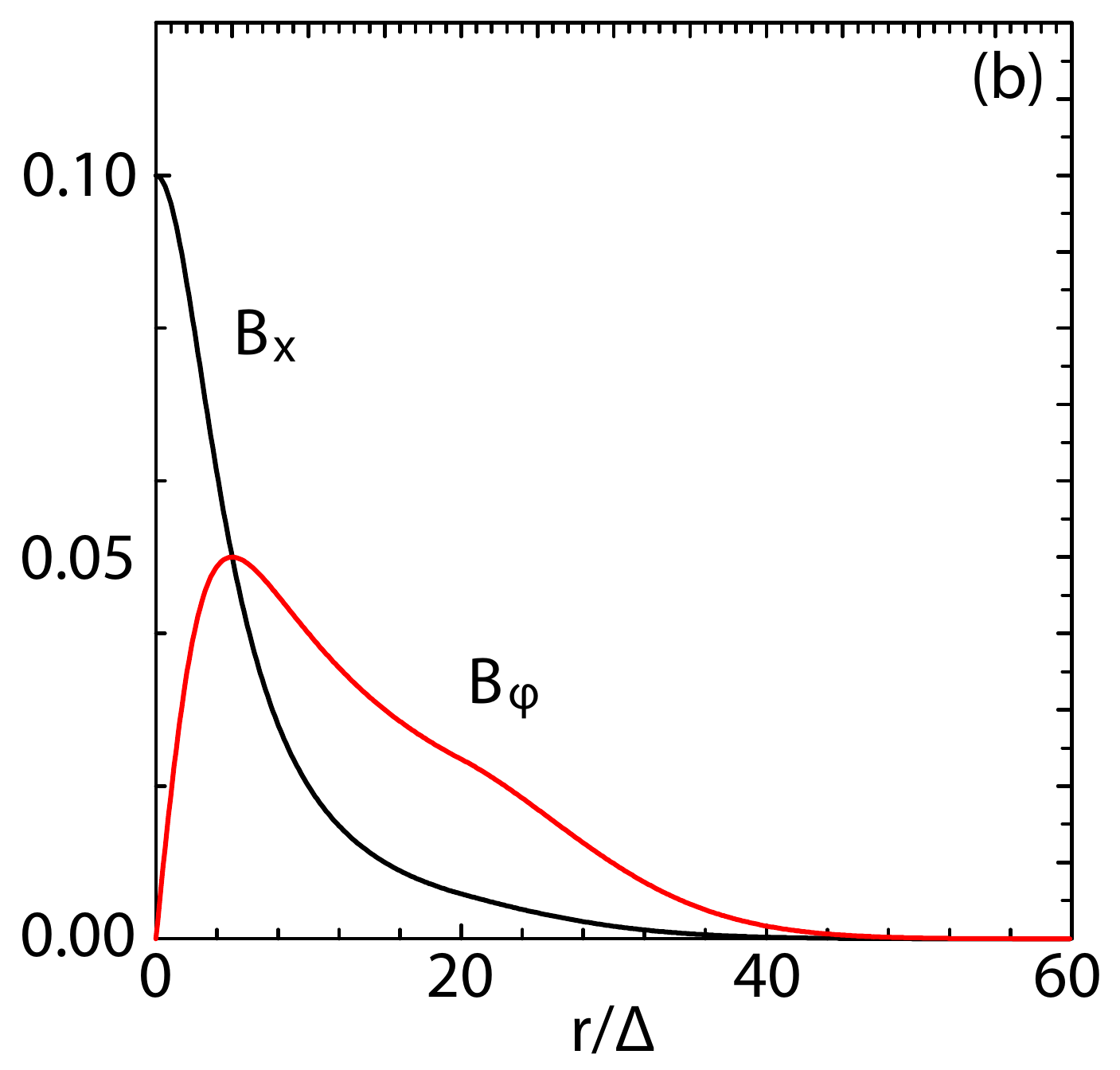}
\caption{Panel (\textbf{a}) shows the schematic global jet simulation setup used in this simulation study.
The~jet is injected at $x = 100\Delta$ with jet radius $r_{\rm jt} = 20\Delta$ at the center of  {the} $y$--$z$
~plane
(not scaled; adapted from \cite{nishi16}). Panel (\textbf{b}) shows the magnetic field component profiles across the jet. The field structure is defined by Equations (1) and (2), with damping applied outside of the jet with length-scale $b = 200.0$ (see text). The jet boundary is located at $r_{\rm jet}=20\Delta$.  The size of the simulation box is small in these simulations {, which}
~are meant to investigate an initial stage of the system evolution.}
\label{hmfg1}
\end{figure}

As an initial step, we have examined how the helical magnetic field modifies jet evolution using a~small system before performing larger-scale simulations, and a schematic of the simulation injection setup  is shown in Figure \ref{hmfg1}a.
\vspace*{-0.0cm}
\textls[-10]{In these small system simulations, we utilize a numerical grid with \mbox{$(L_{x}, L_{y}, L_{z}) = (645\Delta, 131\Delta, 131\Delta)$}} (simulation cell size: $\Delta = 1$) and periodic boundary conditions in transverse directions.  The jet and ambient (electron) plasma number  density measured in the simulation frame is $n_{\rm jt}= 8$ and  $n_{\rm am} = 12$,  respectively.  The cylindrical jet with jet radius $r_{\rm jt} =20\Delta$ is  injected in the middle of   {the} $y$--$z$
~plane ($(y_{\rm jc}, z_{\rm jc}) = (63\Delta, 63\Delta)$) at $x= 100\Delta$.

In the simulations, initial magnetic field amplitude parameter $B_{0}=0.1c$,  {(}$c=1$ {)},
~\mbox{ {(}$\sigma = B^2/n_{\rm e}m_{\rm e}\gamma_{\rm jet}c^{2} =2.8\times 10^{-3}$ {)}}, and $a = 5.0\Delta = 0.25*r_{\rm jt}$ ($r_{\rm jt} = 20\Delta$). The helical field structure inside the jet is defined by Equations (1) and (2). For the external magnetic fields, we use a damping function $\exp{[-(r-r_{\rm jt})^{2}/b]}$ $(r \ge r_{\rm jt})$ that multiplies Equations (1) and (2) with the tapering parameter $b=200$. The final profile of the helical magnetic field components is shown in Figure \ref{hmfg1}b.

In the simulations, the electron skin depth  $\lambda_{\rm s} =  c/\omega_{\rm pe} = 10.0\Delta$, where  $c$ is the speed of light,  $\omega_{\rm pe} = (e^{2}n_{\rm am}
/\epsilon_0 m_{\rm e})^{1/2}$ is the electron plasma frequency, and  the electron Debye length for the ambient
electrons  {is} $\lambda_{\rm D}=0.5\Delta$.  The jet--electron thermal velocity is $v_{\rm jt,th,e} = 0.014c$
in the jet reference frame.  The electron thermal velocity in the ambient plasma is
$v_{\rm am,th,e} = 0.03c$, and  ion thermal velocities are smaller by $(m_{\rm i}/m_{\rm e})^{1/2}$.
Simulations were performed using an electron--positron ($e^{\pm}$) plasma or
an electron--proton ($e^{-}$ {--}$p^{+}$
~with $m_{\rm p}/m_{\rm e} = 1836$) plasma for the jet Lorentz factor of 15
and with the ambient plasma at rest  ($v_{\rm am}= 0$).

\begin{figure}[H]
\vspace{-6pt}

\subfigure[]{\includegraphics[scale=0.36, angle=-90]{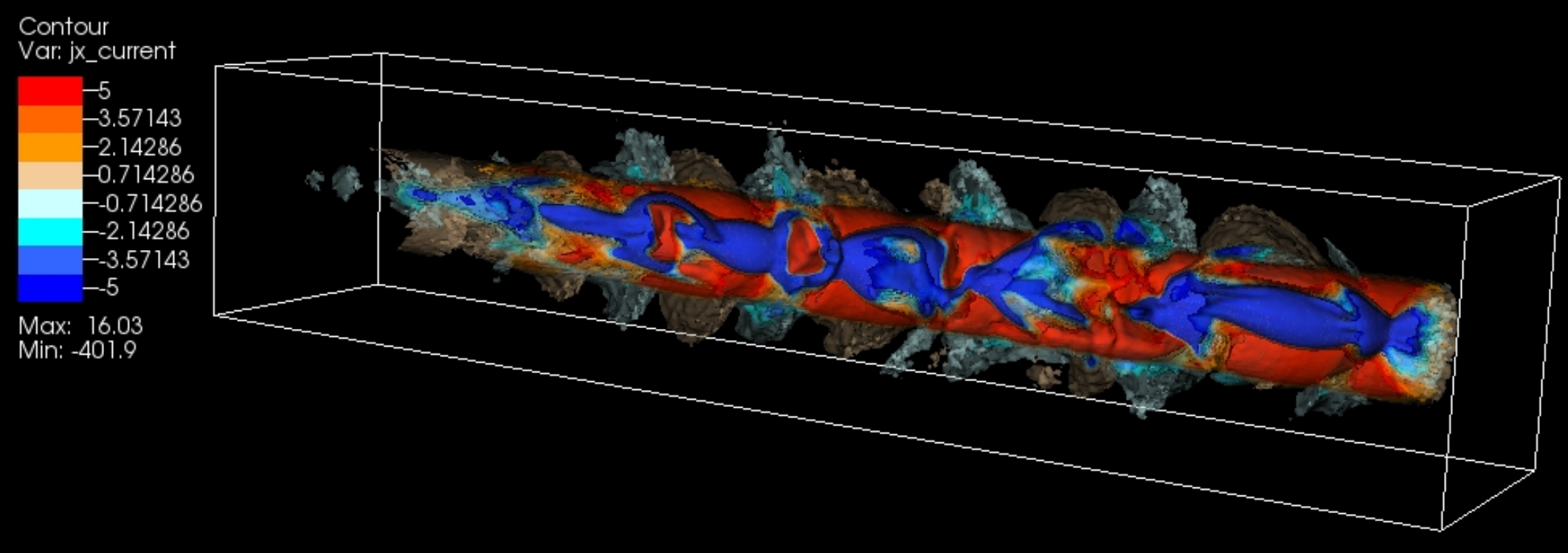}}
\subfigure[]{\includegraphics[scale=0.36, angle=-90]{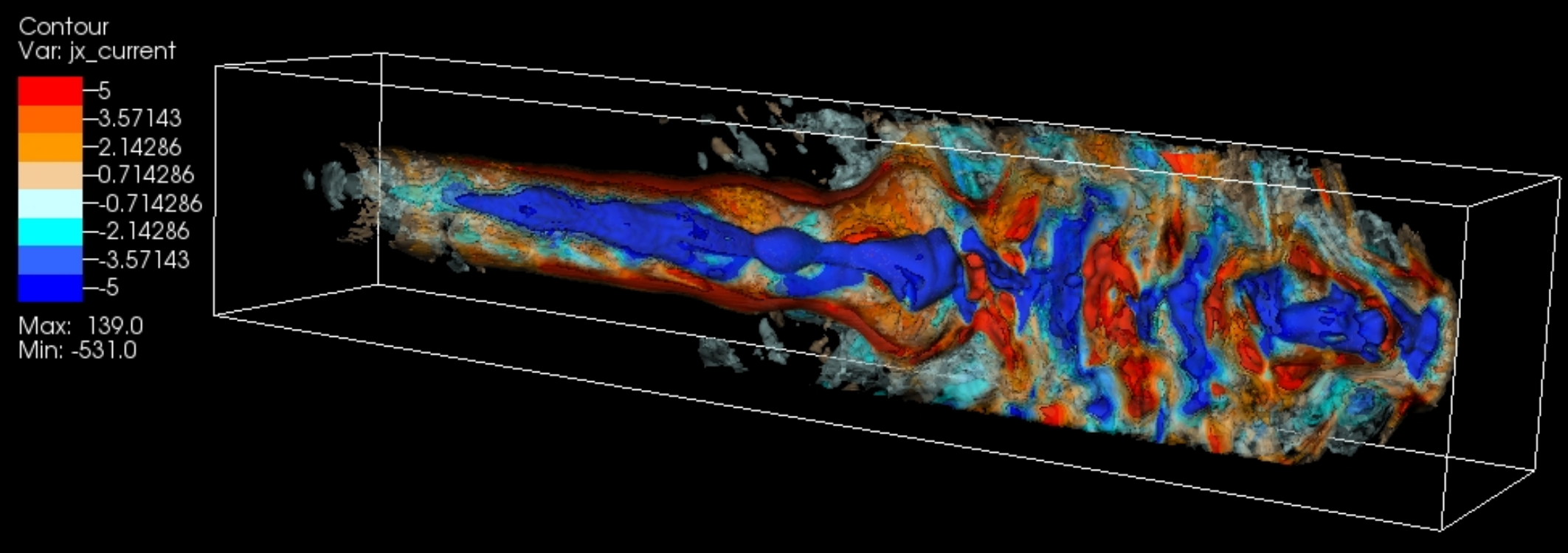}}

\vspace{-12pt}
\caption{Isocontour plots of $J_{\rm x}$ for (\textbf{a}) $e^{-}$ {--}$p^{+}$
; and (\textbf{b})
$e^{\pm}$ jets   at time  $t =  500\,\omega_{\rm pe}^{-1}$. For the $e^{-}$ {--}$p^{+}$
~jet,
recollimation-like shocks  are seen in Figure \ref{2d}a. Figure \ref{2d}b shows the growing instabilities and currents expanding outside the jet, leading to a turbulent current density structure for the $e^{\pm}$ jet.}
\label{2d}
\end{figure}

Figure \ref{2d} shows isocontour plots of the $x$-component of the current density  $J_{\rm x}$ for the   {(a)} $e^{-}$ {--}$p^{+}$
~and (b) $e^{\pm}$ jets   at time
 $t =  500\,\omega_{\rm pe}^{-1}$ ($y/\Delta = 63$).  For  {the} $e^{-}$ {--}$p^{+}$
 ~jet, recollimation-like shocks can be observed
(Figure \ref{2d}a). %
\vspace{-6pt}
The negative current density (blue) is disrupted by the positive current density (red), which
indicates the occurrence of recollimations, like in RMHD simulations \cite{mizuno15}.
In contrast, for the $e^{\pm}$ jet, small recollimation structures occur,  and after instabilities have grown, the currents expand outside the jet and the current density becomes turbulent.

\begin{figure}[H]
\vspace{-6pt}
\subfigure[]{\includegraphics[scale=0.36, angle=-90]{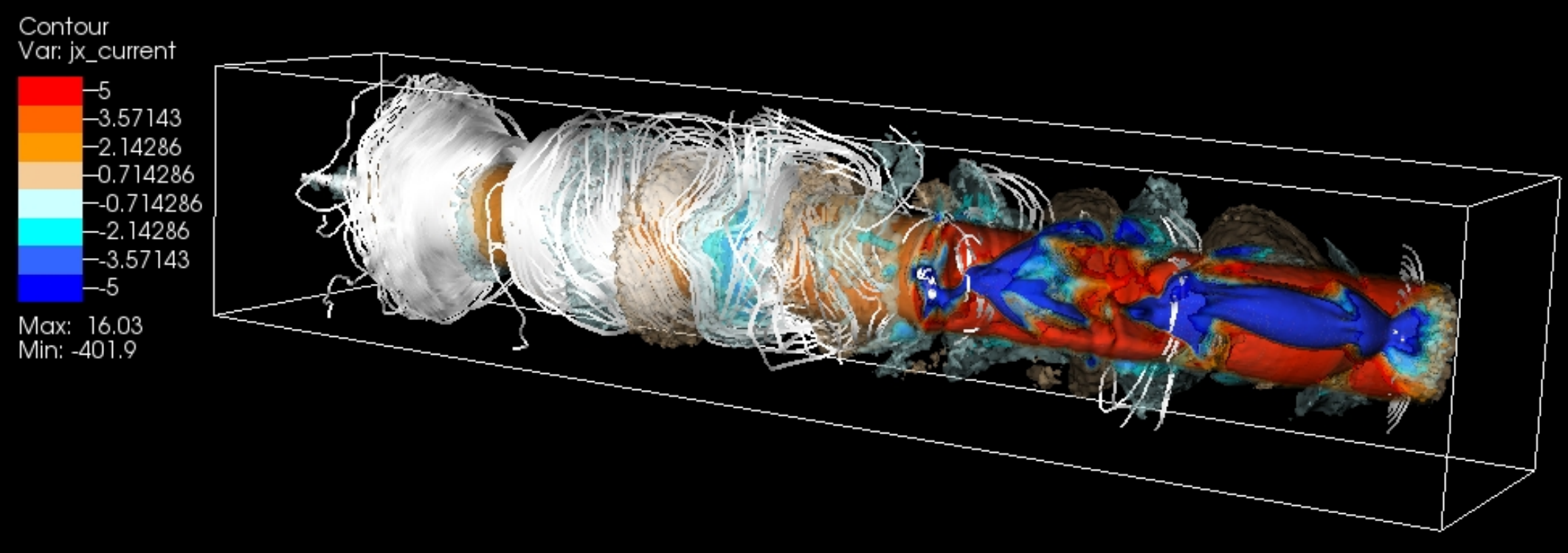}}
\subfigure[]{\includegraphics[scale=0.36, angle=-90]{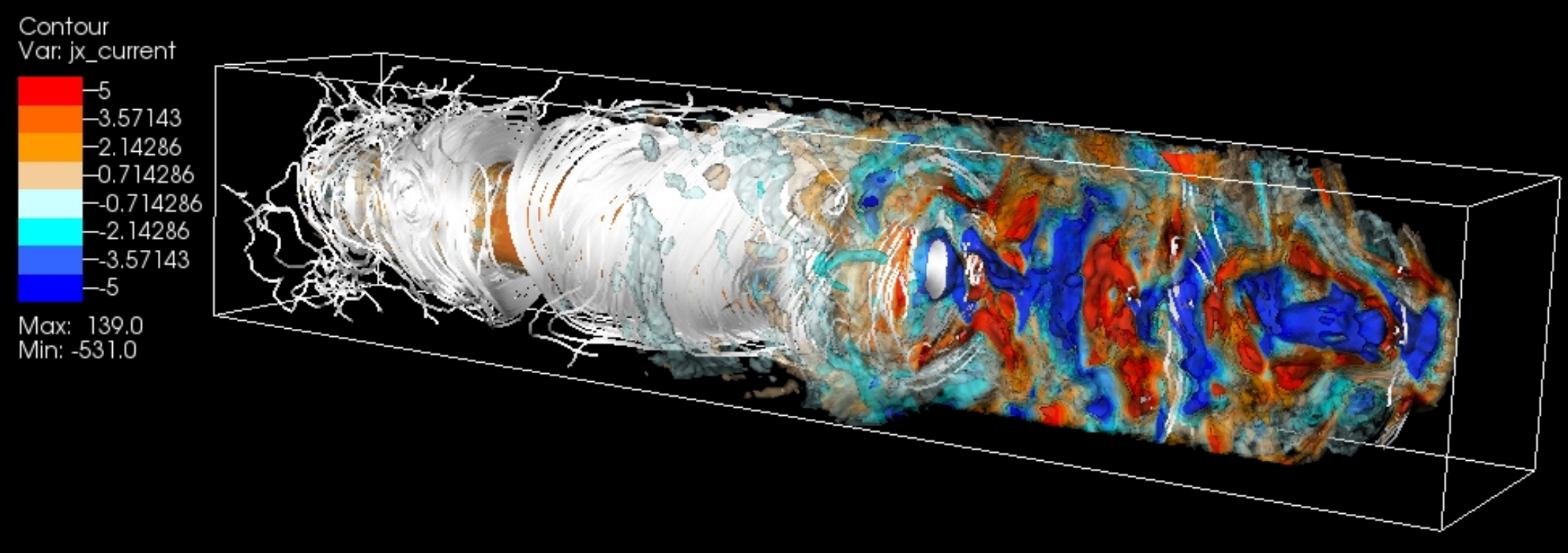}}
\vspace{-12pt}
\caption{\footnotesize{ {Isocontour}
~plots of $J_{\rm x}$  with magnetic field lines  (white lines) for
(\textbf{a}) $e^{-}$ {--}$p^{+}$
; and (\textbf{b}) $e^{\pm}$ jets   at time  $t =  500\,\omega_{\rm pe}^{-1}$. In order to view the inside of
the jets, 3D displays are  clipped along the jets and perpendicular to the jets in the front parts of the figures (compare
Figure \ref{2d})}.}
\label{Bc}
\end{figure}

\begin{figure}[H]

\subfigure[]{\includegraphics[scale=0.43]{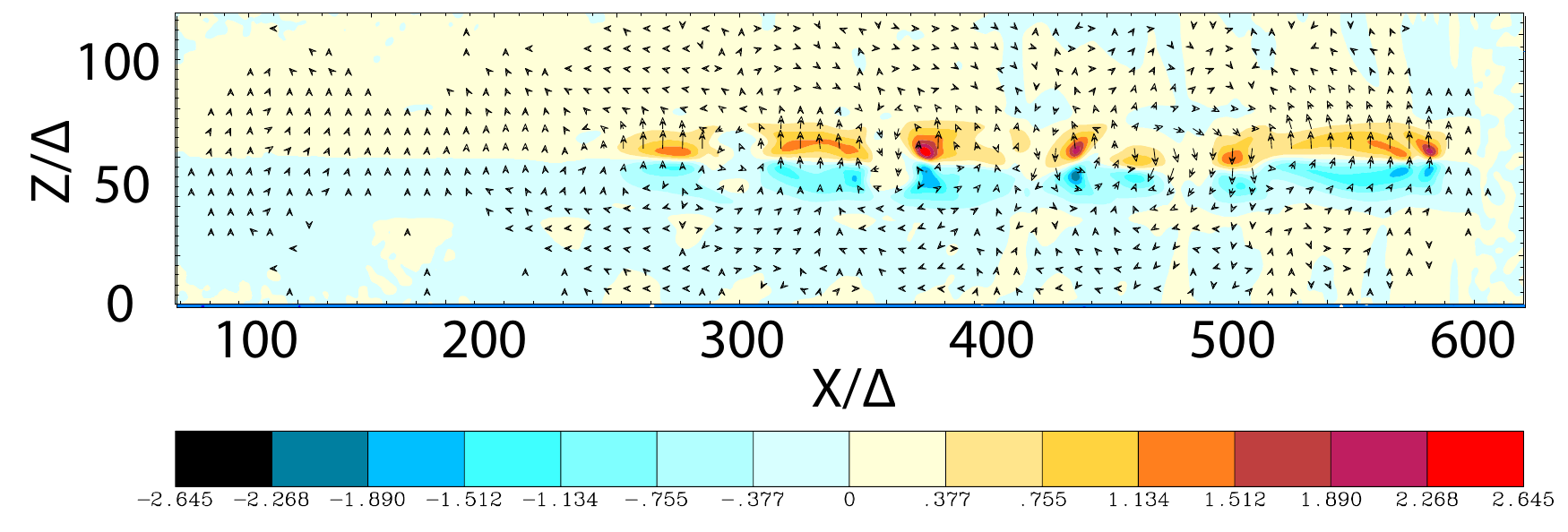}}
\subfigure[]{\includegraphics[scale=0.43]{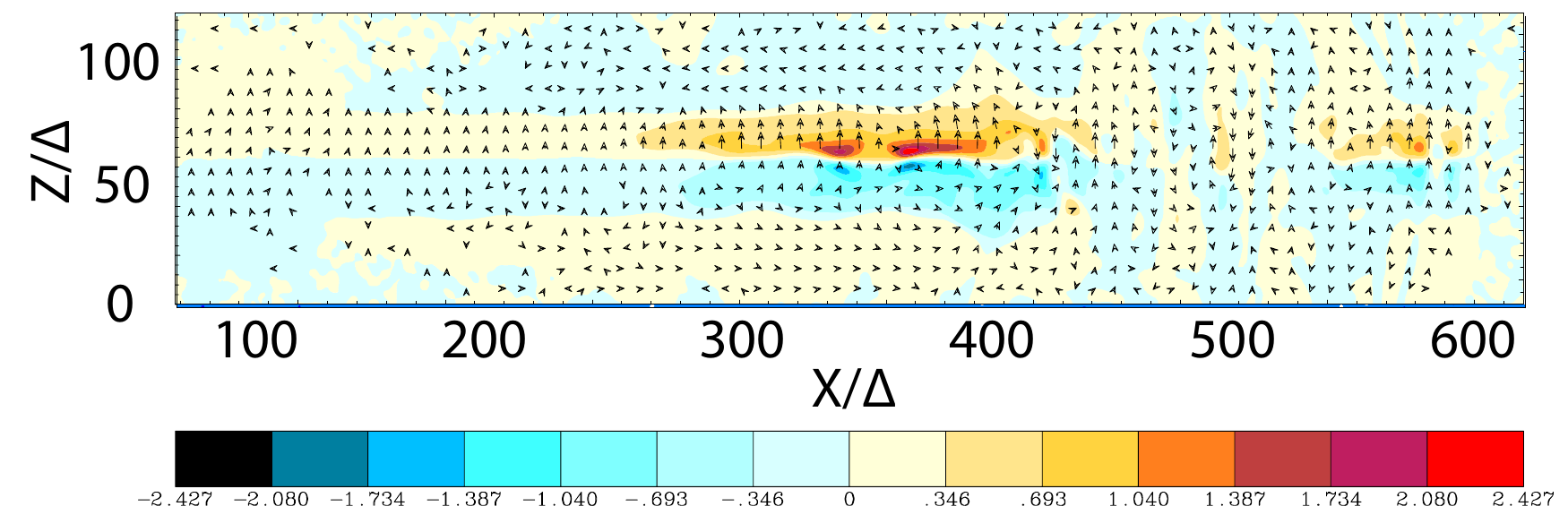}}
\vspace{-0.2cm}
\caption{Isocontour plots of the azimuthal component of magnetic field $B_y$ intensity at the center of the jets
for for (\textbf{a}) $e^{-}$ {--}$p^{+}$
~and (\textbf{b}) $e^{\pm}$ jets  ($y/\Delta = 63$) at time  $t =  500\,\omega_{\rm pe}^{-1}$.  The disruption of
helical magnetic fields  is caused by instabilities and/or  reconnection.}
\label{ByBxz}
\end{figure}

Figure \ref{Bc} shows  the magnetic field lines (white lines) which are carried by the jets.
The helical magnetic fields are distorted due to the instabilities occurring at the jet boundaries (see  Figure \ref{2d}).
\vspace{-6pt}
Figure \ref{ByBxz} shows the $y$ component of the magnetic field ($B_{y}$).  In both cases, the initial helical magnetic field (left-handed; clockwise viewed from the jet front) is enhanced and disrupted due to the~instabilities.

\vspace{-0.0cm}
\subsection{\bf{Particle Acceleration in Helically Magnetized RPIC Jets}}
\vspace{-0.0cm}

\begin{figure}[H]

\subfigure[]{\includegraphics[scale=0.55]{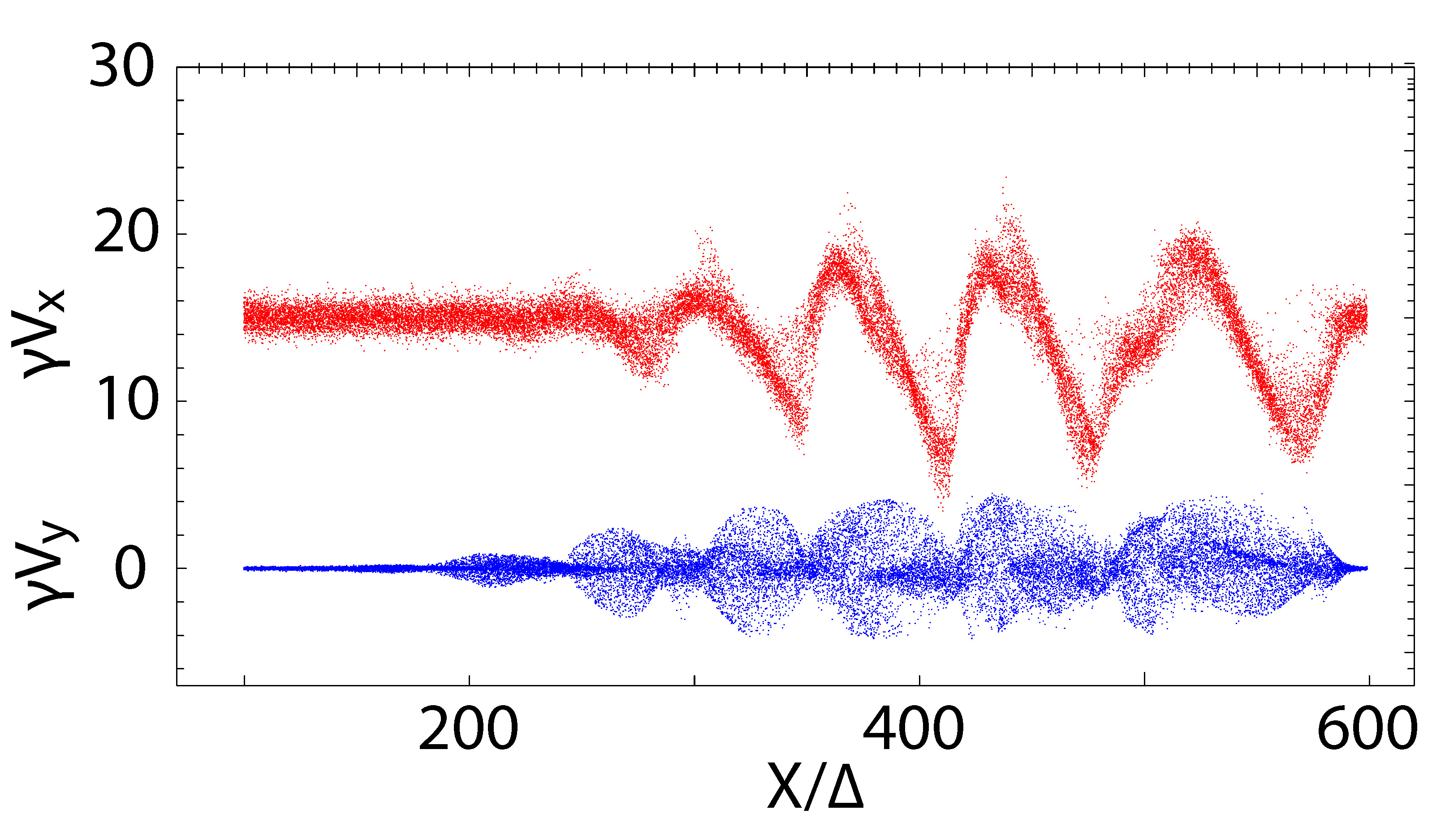}}
\subfigure[]{\includegraphics[scale=0.55]{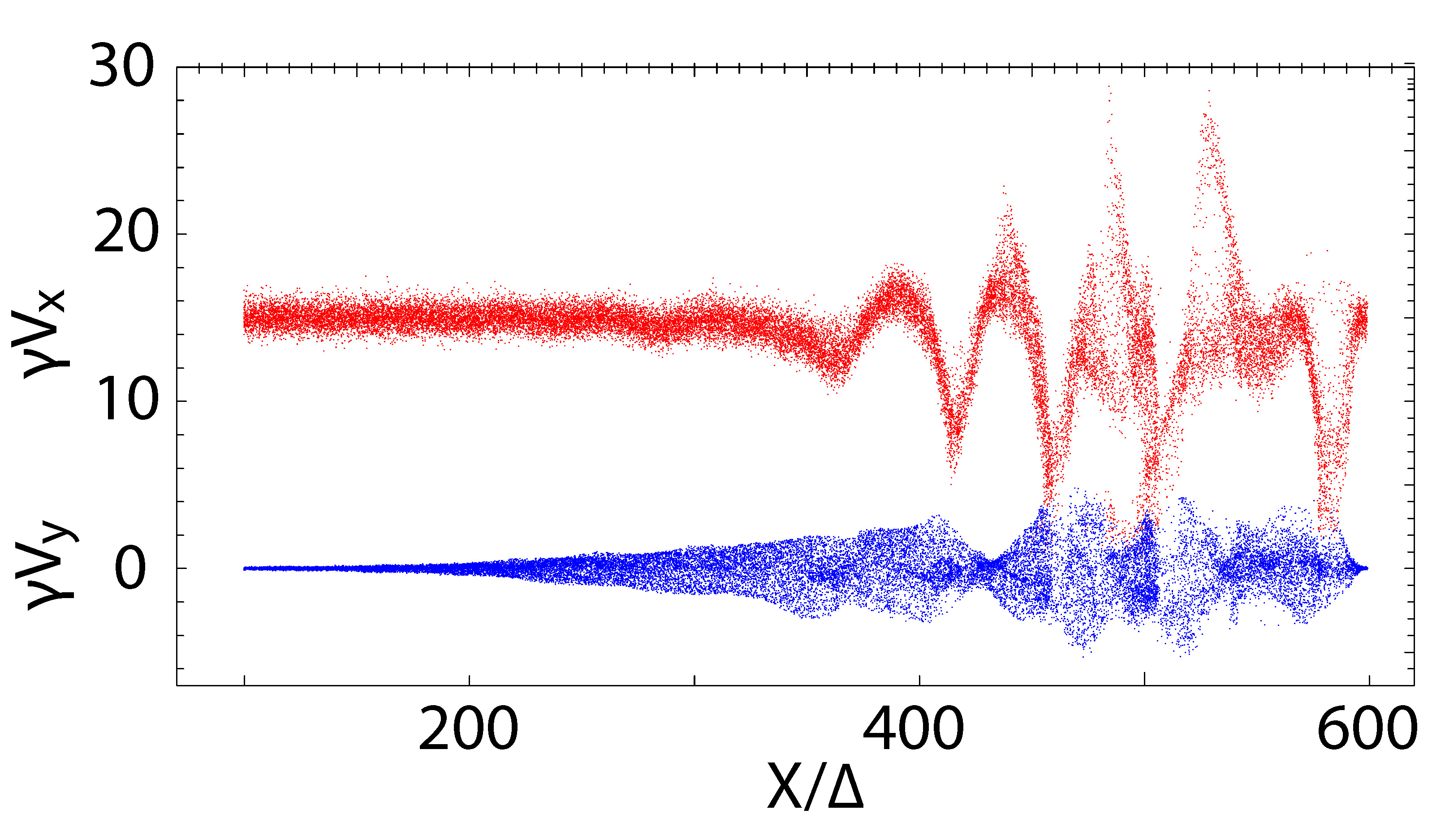}}\\
\vspace{-6pt}
\subfigure[]{\includegraphics[scale=0.55]{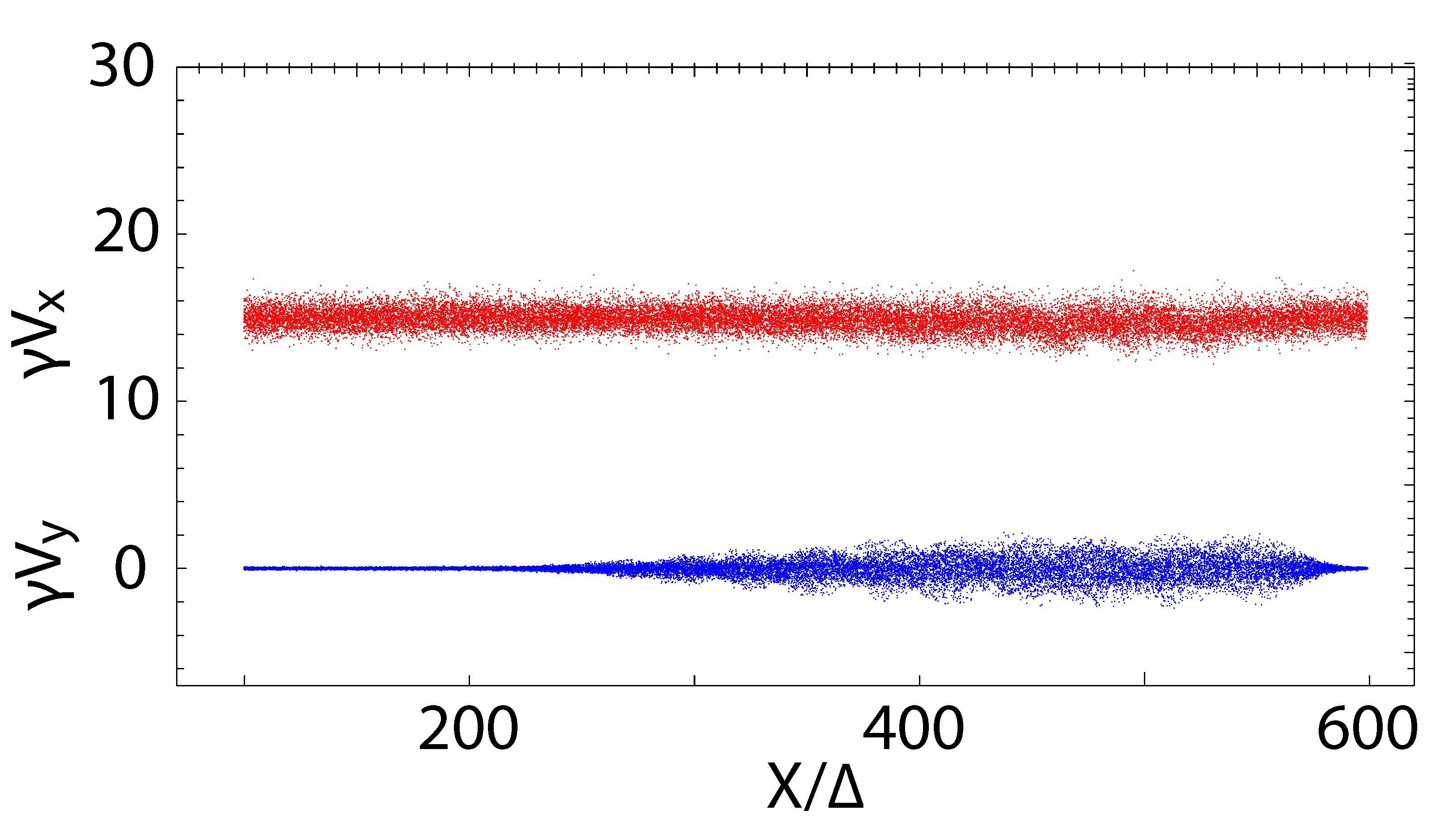}}
\subfigure[]{\includegraphics[scale=0.55]{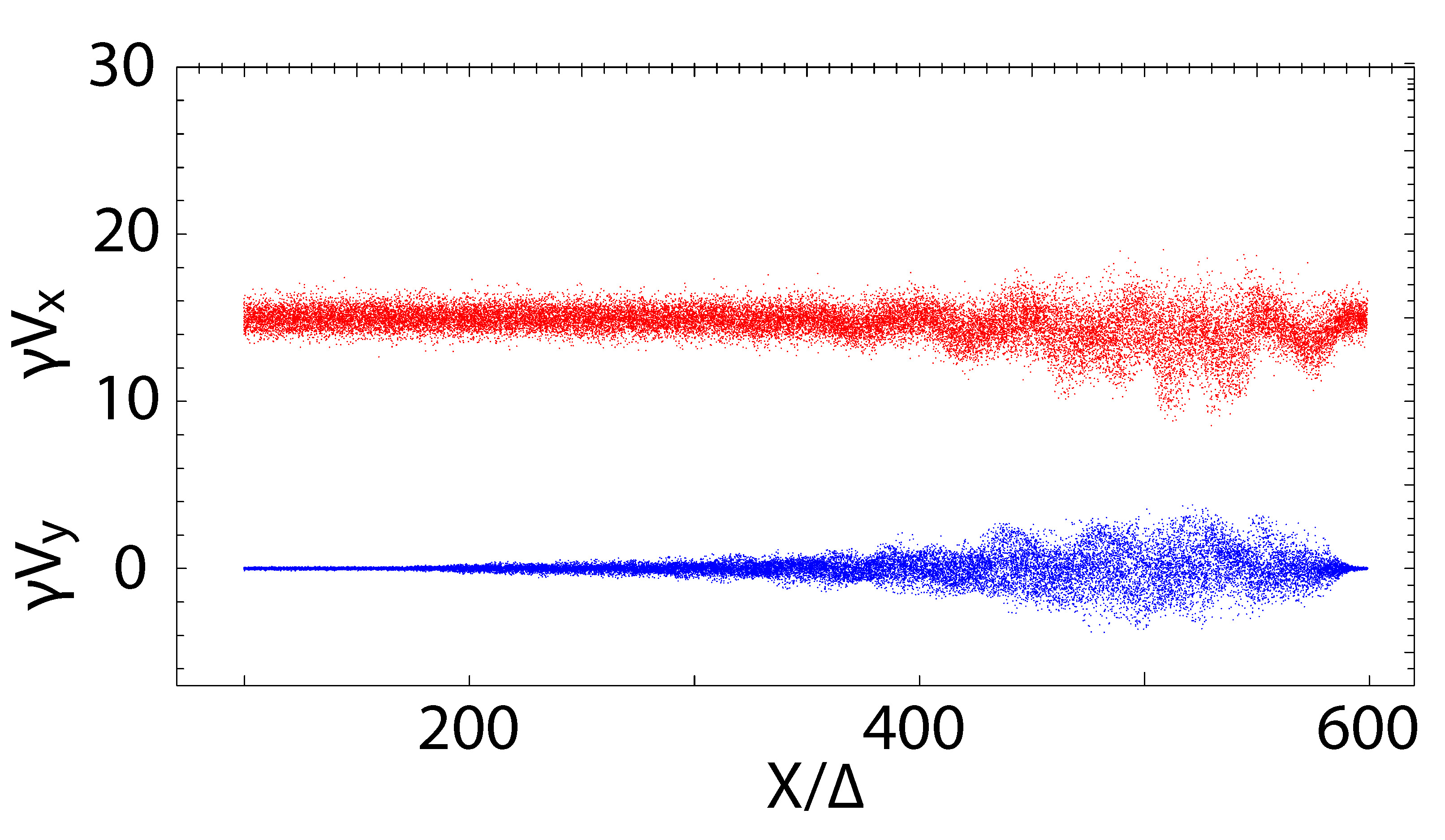}}
\caption{Phase-space plots of jet electrons (red dots: $x-\gamma v_x$, and blue dots:
$x-\gamma v_y$) for (\textbf{a}) $e^{-}$ {--}$p^{+}$
; and (\textbf{b}) $e^{\pm}$ jets at time  $t =  500\omega_{\rm pe}^{-1}$.
Panels (\textbf{c}) and (\textbf{d}) show phase-space plots for simulations without helical magnetic fields.
The strong velocity variation in longitudinal and transverse directions can be caused by recollimation shocks and/or other new instabilities.}
\label{px-v}
\end{figure}

In order to examine particle acceleration, we analyze jet electron phase-space distributions in $x- \gamma v_x, \gamma v_y$.   Figure \ref{px-v} shows phase-space plots for (a)   $e^{-}$ {--}$p^{+}$
~and (b)
$e^{\pm}$ jets  at time $t =  500\,\omega_{\rm pe}^{-1}$. For comparison purposes, the figure includes phase-space plots for simulations without magnetic fields (panels (c) and (d)). The red dots show
$x- \gamma v_x$, and the blue dots show  $x- \gamma v_y$   phase-space distributions. The phase-space distributions indicate strong electron acceleration and deceleration. In the case of the
$e^{-}$ {--}$p^{+}$
~jet (Figure \ref{px-v}a), the acceleration and deceleration occurs both along and transverse to the jet direction. This result is reminiscent of recollimation shocks observed in RMHD simulations \cite{mizuno15}.
On the contrary, in the case of the $e^{\pm}$ jet, the phase-space structure is more complicated and
suggests that reconnection and/or some new instabilities occur.
Further investigation using larger simulation systems is clearly needed to resolve these issues.

\vspace{-0.0cm}
\section{\bf{Comparison to RMHD Results for Jets Containing Helical Magnetic Fields}}
\vspace{-0.0cm}

\begin{figure}[H]
\includegraphics[scale=0.87]{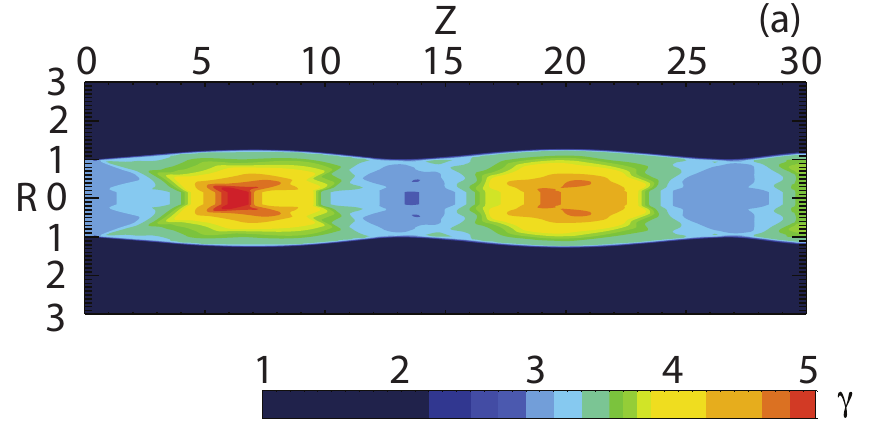}
\hspace*{-0.6cm}
\includegraphics[,scale=0.8]{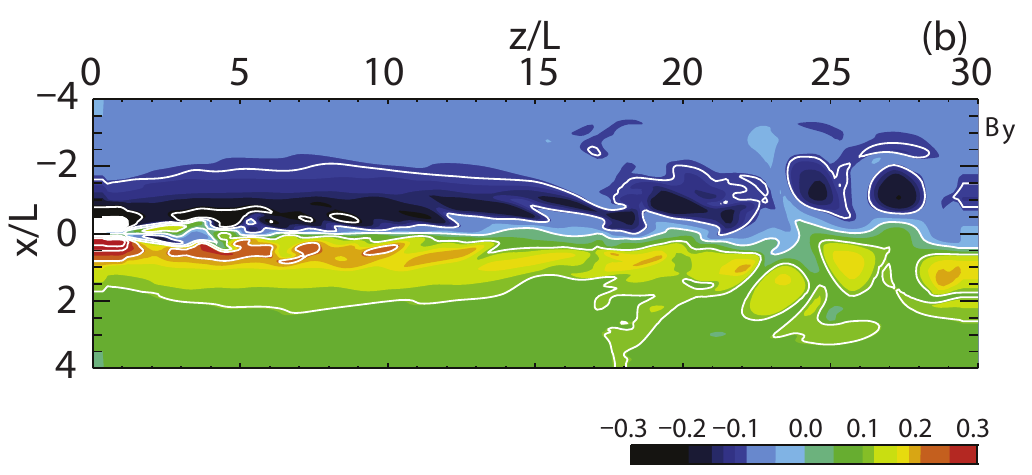}
\includegraphics[,scale=0.43]{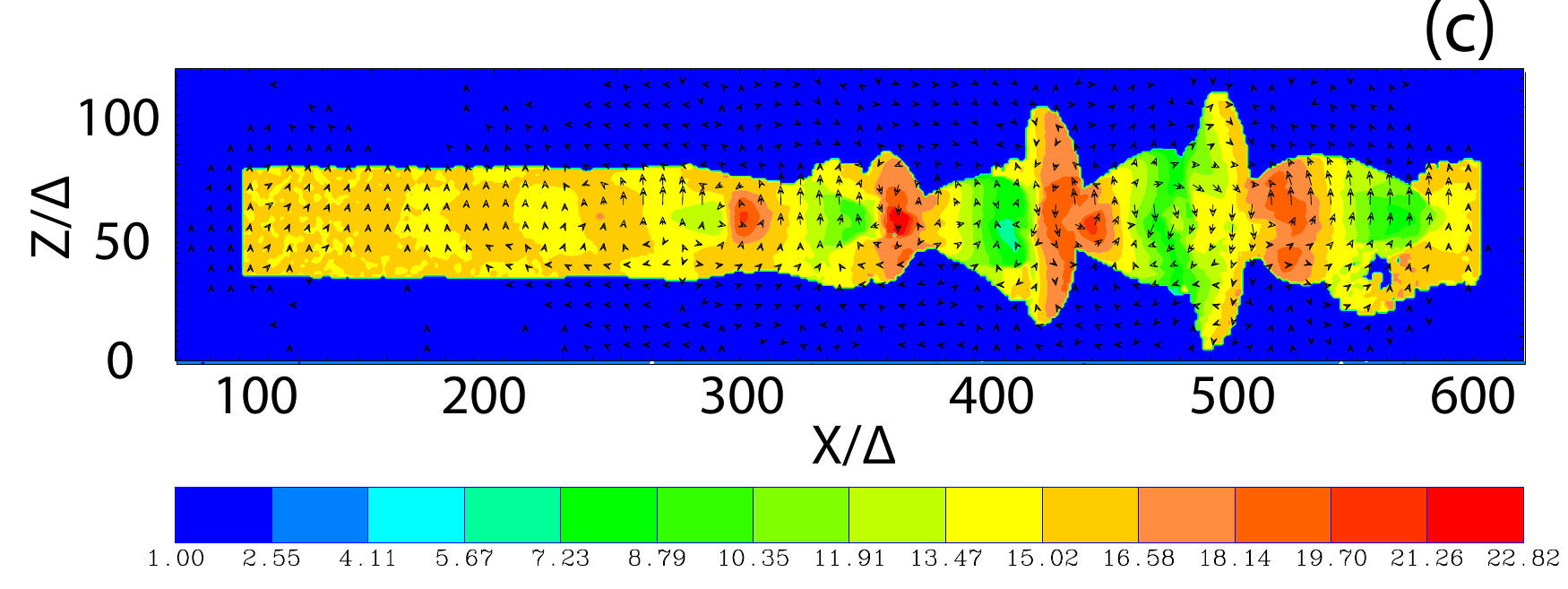}

\caption{Panel (\textbf{a}) shows a 2D plot of the Lorentz factor for a helically-magnetized  {RMHD} jet
~at $t = 200 r_{\rm jt}/c$
~ {(}adapted from Figure 10d in Mizuno et al. \cite{mizuno15} {)};
Panel (\textbf{b}) shows the azimuthal magnetic field component $B_y$ with $|B_y|$ magnitude contours
 for a decreasing density helically-magnetized rotating RMHD jet at $t = 70 r_{\rm jt}/c$. The disruption of helical magnetic fields can be caused by the current-driven kink instability
 ~ {(}adapted from Figure 4f in Singh et al. \cite{singh16} {)}; Panel (\textbf{c}) shows the Lorentz
 factor of jet electrons for $e^{-}$ {--}$p^{+}$
 ~jet ($y/\Delta = 63$)  at time  $t =  500 \omega_{\rm pe}^{-1}$.}
 \label{singhf4f}
\end{figure}

In order to compare our RPIC simulation results with RMHD simulations of (a) recollimation and (b) current-driven
kink instability, two typical RMHD cases are plotted in Figure \ref{singhf4f}.
Figure \ref{singhf4f}a shows the enhanced Lorentz factor due to  recollimation
shocks found by Mizuno et al. \cite{mizuno15}  in two-dimensional special-relativistic magnetohydrodynamic
simulations of non-equilibrium over-pressured helically-magnetized relativistic jets in cylindrical geometry.
Similar structures are observed in our simulations (Figures \ref{2d}a and \ref{ByBxz}a). Figure \ref{singhf4f}c shows
the Lorentz factor of  jet electrons with the observed enhanced Lorentz factor due to the recollimation shocks. As expected from
Figure \ref{px-v}a, the location of the enhanced Lorentz factor corresponds to the location of the jet electrons,
which are accelerated maximally in longitudinal and transverse directions.
However, the evolution of the enhanced Lorentz factor is different compared to the RMHD simulation result.
 The movie of a 2D slice of $J_x$ like Figure \ref{2d}a shows that the recollimation shocks are moving
along the jet, keeping the shape.

\vspace{-0.2cm}
\vspace{-6pt}
Recently, Singh et al. \cite{singh16} showed the spatial development of the current-driven kink instability along
magnetized rotating relativistic jets. Figure \ref{singhf4f}b shows the azimuthal magnetic field component
$B_y$ with $|B_y|$ magnitude contours with right-hand polarity of helical magnetic field (counter-clockwise,
viewed from the jet front). Therefore, the direction of $B_y$ is reversed. The current-driven kink instability
breaks simple helical magnetic fields. This structure in $B_y$ is similar to
that of the $e^{\pm}$ jet case, as shown in Figure \ref{ByBxz}b.



\section{Discussion}


Our initial global jet simulations containing helical magnetic fields show new types of growing instabilities for both
electron--proton and pair plasma jets.  Preliminary results indicate that the presence of helical fields suppresses
the growth of the kinetic instabilities, such as the Weibel instability, kKHI, and MI.
Instead, new instabilities appear, associated with recollimation shocks and  current-driven kink instability.

The $e^{-}$ {--}$p^{+}$
~helically magnetized jet shows recollimation-like shock structures in the current density $J_{x}$,
similar to recollimation shocks observed in RMHD simulations containing helical magnetic fields \cite{mizuno15}.
The observed modulations in the kinetic energy of  jet electrons shown in
Figure \ref{px-v}a  might correspond to the modulations in the Lorentz factor reported in RMHD studies
(see Figure~\ref{singhf4f}a). Additionally, while not shown here, the electron density in the $e^{-}$ {--}$p^{+}$
~jet shows
pile-ups which correspond to recollimation shock structures seen in RMHD simulations. Evidence for  {the} growth of a~kink-like instability in the $e^{\pm}$  jet is seen in the $y$-component
of magnetic field $B_{y}$ in Figure \ref{ByBxz}b, and is similar to that seen in Figure \ref{singhf4f}b, where
helical magnetic fields carried by the jet are disrupted by the growth of the kink instability.

Finally, we see evidence that reconnection is taking place in the jets. However, larger-scale and higher-resolution simulations
are required to fully resolve the reconnection phenomena and to understand the nature of the new
instabilities. Future simulations will be combined with calculations of radiation signatures and polarity, along with variations
in space and time \cite{Zhang15,Zhang16}.







\vspace{6pt}

\supplementary{The following are available online at www.mdpi.com/link,  Video S1: Evolution of $J_{x}$ for the $e^{-}-p^{+}$ jet.}%

\acknowledgments{This work is supported by NSF AST-0908010, AST-0908040,
NASA-NNX09AD16G, NNX12AH06G, NNX13AP-21G,
and NNX13AP14G grants. The work of J.N. and O.K. has been
supported by Narodowe Centrum Nauki through research
project DEC-2013/10/E/ST9/00662.
Y.M. is supported by the ERC Synergy Grant ``BlackHoleCam - Imaging
the Event Horizon of Black Holes''  (Grant No. 610058).  M.P.~acknowledges
support through grant PO 1508/1-2 of the Deutsche Forschungsgemeinschaft.
Simulations were performed using  Pleiades and Endeavor facilities at NASA
Advanced Supercomputing (NAS), and using Gordon and Comet at The San Diego
Supercomputer Center (SDSC), and Stampede at The Texas Advanced
Computing Center, which are supported by the NSF. This research was started during the program
``Chirps, Mergers and Explosions: The Final Moments of Coalescing Compact Binaries''
at the Kavli Institute for Theoretical Physics, which is supported by the National Science
Foundation under grant No. PHY05-51164. The first velocity shear results using
an electron$-$positron plasma were obtained during the Summer Aspen workshop
``Astrophysical Mechanisms of Particle Acceleration and Escape from the Accelerators''
held at the Aspen Center for Physics (1--15 September 2013).}

\authorcontributions{{
K. -I. Nishikawa: Perform simulations, analyze the data and prepare a manuscript; Y. Mizuno: Compare with RMHD simulations; J. Niemiec: Contribute modifying the code for this research; O. Kobzar; Modify the code for this simulation; M. Pohl: Overlook the simulation results; J. L. G\'{o}mez : Contribute on comparisons with observations; I. Du\c{t}an: Perform some of simulations for this research; A. Pe'er: Critical contributions for physical interpretation; J. T. Frederiksen: Contribution for critical discussions on this research; \AA. Nordlund: Fruitful suggestions for this research; A. Meli: Critical reading and discussion on this research; H. Sol: Essential suggestions for this research; P. E. Hardee: Theoretical contributions for this research; D. H. Hartmann: Useful discussions for this research}}

\conflictofinterests{
The authors declare no conflict of interest.
}



\appendixtitles{no} 


\bibliographystyle{mdpi}

\renewcommand\bibname{References}


\begin{thebibliography}{999}
\bibitem{peer14}
Pe'er, A. Energetic and Broad Band Spectral Distribution of Emission from Astronomical Jets.
{\em Space Sci. Rev.} {\bf 2014}, {\em 183}, 371--403.
\bibitem[Blandford \& Znajek(1977)]{blanz77}
Blandford, R.D.;  Znajek, R.L.
Electromagnetic extraction of energy from Kerr black holes.
{\em Mon. Not. R.  Astron. Soc.} {\bf 1977}, {\em 179}, 433--456.
\bibitem{mck14}
McKinney, J.C.; Tchekhovskoy, A.;   Sadowski, A.; Narayan, N.
Three-dimensional general relativistic radiation magnetohydrodynamical
simulation of super-Eddington accretion, using a new code HARMRAD
with M1 closure.
{\em Mon. Not. R.  Astron. Soc.} {\bf 2014}, {\em 441}, 3177--3208.
\bibitem[Laing(1981)]{laing81} 	
Laing, R.A.	
Magnetic fields in extragalactic radio sources.
{\em Astrophys. J.} {\bf 1981}, {\em 248}, 87--104.
\bibitem[Aloy et al.(2000)]{aloy00}
Aloy, M.; G\'{o}mez, J.; Ib\'{a}\~nez, J.M.; Mart\'i, J.M.; M\"uller, E.
Radio Emission from Three-dimensional Relativistic Hydrodynamic Jets: Observational Evidence of Jet Stratification.
{\em Astrophys. J.} {\bf 2000}, {\em 528}, L85--L88.
\bibitem[Clausen-Brown, Lyutikov \& Kharb(2011)]{Clausen11}
Clausen-Brown, E.; Lyutikov, M.;   Kharb, P.
Signatures of large-scale magnetic fields in active galactic nuclei jets:
Transverse asymmetries.
{\em Mon. Not. R.  Astron. Soc.} {\bf 2011}, {\em 415}, 2081--2092.
\bibitem[Meier(2008)]{meier08}
Meier, D.L. Astrophysics: Exhaust inspection. {\em Nature} {\bf 2008}, {\em 452}, 945--946.
\bibitem[Piran(2004)]{piran04} Piran, T. The physics of gamma-ray bursts. {\em Rev. Mod. Phys.}
{\bf 2004}, {\em76}, 1143--1214.
\bibitem[O'Sullivan et al.(2013)]{OS13}
O'Sullivan, S.P.; McClure-Griffiths, N.M.; Feain, I.J.;  Gaensler, B.M.;
Saul, R.J.
Broad-band radio circular polarization spectrum of the relativistic jet
in PKS B2126-158.
{\em Mon. Not. R.  Astron. Soc.} {\bf 2013}, {\em 435}, 311--319.
\bibitem[Wardle et al.(19988)]{wardle98}
Wardle, J.F.C.; Homan, D.C.; Ojha, R.;  Roberts, D.H.
Electron-positron jets associated with the quasar 3C279.
{\em Nature} {\bf 1998}, {\em 395}, 457--461.
\bibitem[Buneman(1993)]{buneman93}
Buneman, O.   {\em Computer Space Plasma Physics:
Simulation Techniques and Software};  Matsumoto, H.,
 Omura,~Y.,~Eds.; Terra Scientific Publishing Company: Tokyo, Japan, 1993; pp. 67--79.
\bibitem[Niemiec et al.(2008)]{niem08}
Niemiec, J.;  Pohl, M.; Stroman, T.; Nishikawa, K.-I.   Production of Magnetic Turbulence by Cosmic Rays Drifting Upstream of Supernova Remnant Shocks. {\em Astrophys. J.} {\bf 2008}, {\em 684}, 1174--1189.
\bibitem[Nishikawa et al.(2009)]{nishi09}
Nishikawa, K.-I.;  Niemiec, J.;  Medvedev, M.;   Sol, H.;  Hardee, P.;   Mizuno, Y.;  Zhang, B.;  Pohl, M.;  Oka, M.;
 Hartmann, D.H. Weibel instability and associated strong fields in a fully 3D simulation of a relativistic shock.
{\em Astrophys. J. Lett.} {\bf 2009}, {\em 698}, L10--L14.
\bibitem[Nishikawa et al.(2016)]{nishi16}
Nishikawa, K.-I.;  Frederiksen, J.T.;   Nordlund, \AA.;  Mizuno, Y.;  Hardee, P.E.;
 Niemiec, J.;   G\'{o}mez, J.L.;   Pe'er,  A.;  Du\c{t}an, I.;    Meli, A.;   et al.
Evolution of Global Relativistic Jets:
Collimations and Expansion with kKHI and the Weibel Instability.
{\em Astrophys. J.} {\bf 2016},  {\em 820}, 94--107.
\bibitem[Alves et al.(2012)]{Alves12}
Alves, E.P.; Grismayer, T.; Martins, S.F.; Fi\'{u}za, F.; Fonseca, R.A.; Silva, L.O.
Large-scale magnetic field generation via the Kelvin-Helmholtz instability in unmagnetized scenarios. {\em Astrophys. J. Lett.}
{\bf 2012}, {\em  746}, L14--L19.    
\bibitem[Alves et al.(2014)]{Alves14}
Alves, E.P.; Grismayer, T.;  Fonseca, R.A.; Silva, L.O.
Electron-scale shear instabilities: Magnetic field generation and particle acceleration in
astrophysical jets. {\em New J. Phys.} {\bf 2014}, {\em 16},  {035007}.   
\bibitem[Alves et al.(2015)]{Alves15}
Alves, E.P.; Grismayer, T.;  Fonseca, R.A.; Silva, L.O.
Transverse electron-scale instability in relativistic shear flows.
{\em Phys. Rev. E} {\bf 2015}, {\em 92}, 021101.
\bibitem[Grismayer et al.(2013a)]{Gris13a}
Grismayer, T.; Alves, E.P.;  Fonseca, R.A.; Silva, L.O.
dc-Magnetic-Field Generation in Unmagnetized Shear Flows. {\em Phys. Rev. Lett.} {\bf 2013}, {\em 111}, 015005.
\bibitem[Grismayer et al.(2013b)]{Gris13b}
Grismayer, T.; Alves, E.P.;  Fonseca, R.A. Silva, L.O.
Theory of multidimensional electron-scale instabilities in unmagnetized shear flows.
{\em Plasma Phys. Controll. Fusion} {\bf 2013b}, {\em 55}, 124031.
\bibitem[Nishikawa et al.(2013a)]{nishi13a}
Nishikawa, K.-I.;  Hardee, P.;  Mizuno, Y.;  Dutan,  I.;  Zhang,  B.;  Medvedev, M.;  Choi, E.J.;   Min, K.W.;
Niemiec,~J.;  Nordlund, \AA.;  et al. Radiation from relativistic jets from particles accelerated by shocks, shear-flows, and~reconnection. In Proceedings of the EPJ Web of Conferences,   {Granada, Spain, 10--14 June  2013}; Volume~61,  {02003}.

\bibitem[Nishikawa et al.(2013b)]{nishi13b}
Nishikawa, K.-I.;    Zhang, B.;  Dutan, I.;  Medvedev, M.;  Hardee,  P.;  Choi, E.-J.;  Min, K.;
Niemiec, J.;  Mizuno,~Y.;   Nordlund, \AA.;  et al.
 Magnetic field generation in a jet-sheath plasma via the
kinetic Kelvin-Helmholtz instability. {\em Ann. Geophys.} {\bf 2013}, {\em 31}, 1535--1541.
\bibitem[Nishikawa et al.(2014a)]{nishi14a}
Nishikawa, K.-I.;  Hardee,  P.;   Dutan, I.;
Niemiec,  J.;  Medvedev, M.;  Mizuno, Y.;   Meli, A.;   Sol, H.;   Zhang,  B.;   Pohl, M.; et al. Magnetic field generation via the kinetic Kelvin-Helmholtz instability in core-sheath jets.
{\em Astrophys. J.} {\bf 2014}, {\em 793}, 163--248. 

\bibitem[Liang et al.(2013a)]{liang13a}
Liang, E.; B\"{o}ttcher, M.; Smith, I. Magnetic Field Generation and Particle Energization at
Relativistic Shear Boundaries in Collisionless Electron-Positron Plasmas. {\em Astrophys. J. Lett.} {\bf 2013},
{\em 766}, L19--L24.
\bibitem[Liang et al.(2013b)]{liang13b}
Liang, E.; Fu, W.; B\"{o}ttcher, M.;  Smith, I.; Roustazadeh, P. Relativistic Positron-Electron-Ion Shear Flows
and Application to Gamma-Ray Bursts. {\em Astrophys. J. Lett.} {\bf 2013}, {\em 779}, L27--L34.

\bibitem[Alves(2010)]{Alves10}
Alves, E.P. Magnetic Field Generation via the Kelvin-Helmholtz Instability. Master's Thesis,  {Universidate~T\'{e}mica de Lisbon},  2010.
\bibitem[Nishikawa et al.(2014b)]{nishi14b}
Nishikawa, K.-I.; Hardee, P.;  Dutan,  I.;   Zhang, B.;   Meli, A.;   Chjoi, E.-J.; Min,   K.;
Niemiec, J.;  Mizuno,~Y.;  Medvedev,  M.;   et al. Radiation from Particles Accelerated in Relativistic Jet Shocks and
Shear-fows. In~Proceedings of the  {2014 Fifth International Fermi Symposium eConf C14102.1},  {Nagoya, Japan, 20--24~October  2014}.

\bibitem[Nishikawa et al.(2003)]{nishi03}
Nishikawa, K.-I.; Hardee,
P.; Richardson, G.; Preece, R.; Sol, H.; Fishman, G.J. Particle Acceleration in Relativistic Jets Due to Weibel Instability. {\em Astrophys. J.} {\bf 2003}, {\em 595}, 555--563.
\bibitem[Nishikawa et al.(2005)]{nishi05}
Nishikawa, K.-I.; Hardee,
P.; Richardson, G.; Preece, R.; Sol, H.;  Fishman, G.J.
 Particle Acceleration and Magnetic Field Generation in Electron-Positron Relativistic Shocks. {\em Astrophys. J.} {\bf 2005},
 {\em 622}, 927--937.
 \bibitem[Ng \& Noble(2006)]{ng06}
Ng, J.S.T.;  Noble, R.J.
Inductive and Electrostatic Acceleration in Relativistic Jet-Plasma Interactions. \mbox{{\em Phys. Rev. Lett.}} {\bf 2006}, {\em 96},
115006.

\bibitem[Mizuno et al.(2014b)]{mizuno14}  
Mizuno, Y.;  Hardee, P.E.;  Nishikawa, K.-I.
Spatial Growth of the Current-Driven Instability in Relativistic Jets.
{\em  Astrophys. J.} {\bf 2014}, {\em 784}, 167.
\bibitem[Singh et al.(2016)]{singh16}
Singh, C.B.;  Mizuno, Y.;   de Gouveia Dal Pino, E.M.
Spatial Growth of Current-driven Instability in Relativistic Rotating Jets and the Search for Magnetic Reconnection.
{\em Astrophys. J.} {\bf 2016}, {\em 824}, 48.

\bibitem[Markidis et al.(2014)]{markid14}
Markidis, S.; Lapenta, G.;   Delzanno, G.L.;   Henri, P.;   Goldman, M.V.;
Newman, D.L.;  Intrator, T.;    Laure,~E.
Signatures of secondary collisionless magnetic reconnection driven by kink
instability of a flux rope. {\em Plasma~Phys. Controll. Fusion} {\bf 2014}, {\em 56}, 064010.
\bibitem[Mizuno et al.(2015)]{mizuno15}  
Mizuno, Y.;  G\'{o}mez, J.L.; Nishikawa, K.-I.;  Meli, A.; Hardee, P.E.; Rezzolla, L.
Recollimation Shocks in Magnetized Relativistic Jets.
{\em Astrophys. J.} {\bf 2015}, {\em 809}, 38.

 \bibitem[Zhang et al.(2015)]{Zhang15}
Zhang, H.; Chen, X.;  B\"{o}ttcher, M.;  Guo, F.;  Li, H.
Polarization Swings Reveal Magnetic Energy Dissipation in Blazars.
{\em Astrophys. J.} {\bf 2015},  {\em 804}, 60--68.
\bibitem[Zhang et al.(2016)]{Zhang16}
Zhang, H.; Deng, W.; Li, H.;  B\"ottcher, M.
Polarization Signatures of Relativistic Magnetohydrodynamic Shocks in the Blazar Emission Region. I. Force-free Helical Magnetic Fields.
{\em Astrophys. J.} {\bf 2016}, {\em 817}, 63. 

\end{thebibliography}


\end{document}